\documentclass[reprint,aps,prb,superscriptaddress,twocolumn,floatfix]{revtex4-2}
\usepackage[english]{babel}
\usepackage{tikz}
\usepackage{graphicx,graphics,grffile}% Include figure files
\usepackage{dcolumn}% Align table columns on decimal
\usepackage[abs]{overpic}
\usepackage[a4paper,top=2cm,bottom=2cm,left=2cm,right=2cm]{geometry}

\usepackage{amsmath,amssymb,amsfonts}
\usepackage{xcolor}
\usepackage[breaklinks=true, colorlinks,citecolor=blue,linkcolor=blue,urlcolor=blue]{hyperref}
\usepackage{braket}
\usepackage{bm}
\usepackage{listings}

\begin{document}

\title{Manipulating non-abelian anyons of Kitaev quantum spin liquids with local magnetic fields}
\author{Haoran Wang}
\email{haoran.wang-6@postgrad.manchester.ac.uk}
\affiliation{\noindent Department of Physics and Astronomy, University of Manchester, Manchester M13 9PL, UK}
\author{Alessandro Principi}
\email{alessandro.principi@manchester.ac.uk}
\affiliation{\noindent Department of Physics and Astronomy, University of Manchester, Manchester M13 9PL, UK}

%TODO: Mention entanglement, entanglement decay times (short paper),  

\begin{abstract}
% Quantum spin liquids hosting non-abelian anyons have recently experienced renewed interest following the discovery of a variety of materials proximate to these quantum phases. Their
Non-abelian anyonic excitations of quantum spin liquids have potential for application to topological quantum computation, but designing logical operations requires developing protocols to faithfully create, move, and read-out such quasiparticles. In this paper, we present a protocol that manipulates the $Z_2$ fluxes (``visons''), and their bound Majorana zero mode, of the Kitaev model. This is achieved by adiabatically switching-off and -on interaction terms in the Hamiltonian, and applying a local magnetic field.
We test our protocol by exchanging two Majorana zero modes, analysing the errors it produces.
We find that the error rate of our protocol can be suppressed exponentially by increasing the adiabatic turn-on time and the size of the system. 
{However, realistic implementations must consider the trade off between different error sources.}
%We also comment on the feasibility of our protocols in solid-state devices and quantum emulators.
%we test two such protocols for manipulating $Z_2$ fluxes (``visons'') of ferromagnetic and antiferromagnetic Kitaev models perturbed by a small uniform magnetic field. Our design employs a driving magnetic field that is locally applied to the sites of the lattice.
%We show that this protocol suffers from significant unavoidable errors that, accumulating over time, prevent reaching high probabilities of generating and displacing flux pairs. However, if control is given on individual spin couplings, then it is possible to create and displace fluxes with nearly $100\%$ probability. We show that braiding non-abelian anyons can be achieved with good precision with our protocol.
%
%
%and can achieve high probabilities of generating and displacing flux pairs of both the $A_z$ and $B$ phases of the model.
%We also analyse the stability of flux pairs find conditions on the external field for fluxes to remain 
\end{abstract}

\maketitle
{\it Introduction}---One of the main challenges to bring about large-scale quantum computation is the decoherence \cite{chuang1995quantum,zurek2003decoherence,hornberger2009introduction,schlosshauer2007decoherence,zurek1991decoherence} of employed quantum states. While error-correction codes \cite{shor1995scheme,jordan2006error,knill1997theory,shor1996fault,cory1998experimental,chiaverini2004realization,knill2001benchmarking} have been designed, they usually require duplication of information over several physical qubits to encode a single logical qubit. This fact hinders scalability of quantum computer architectures. It is thus highly desirable to develop qubits that are by their own design robust against noise.

Topological quantum computation \cite{freedman2003topological,toriccode,dennis2002topological,stern2013topological,das2006topological} offers a potential solution to the decoherence problem. In certain topologically ordered quantum systems, for example in the Kitaev toric code \cite{toriccode}, the ground state manifold of the Hilbert space fractures into different independent sectors \cite{PhysRevB.41.9377}. In order to go from one sector to another, one must apply a chain of operators whose length scales with the dimensions of the system. It follows that a quantum superposition of states in different sectors is, in a macroscopic system, naturally protected from decoherence. It is in fact unlikely for environmental noise to produce a sufficiently long (and non-local) string of operators that can irreparably damage the quantum state of the system. 
Furthermore, local errors can be detected and corrected by measuring local operators named stabilizers. \cite{toriccode}
In this computation scheme, the manipulation of quantum states is achieved by braiding excitations \cite{nayak2008non} which are, in general, anyons.
%({\it i.e.} they behave as neither bosons nor fermions). 
When one such particle is dragged around another, the state of the system is unitarily rotated within the ground state manifold and a quantum gate is effectively applied on it. Thus, quantum algorithms can be elegantly designed by braiding anyonic excitations.

The Kitaev model \cite{Kitaev_2006,hermanns2018physics,jackeli2009mott,takagi2019concept,winter2017models} of a quantum spin liquid, a close relative of the toric code~\cite{Kitaev_2006}, offers a platform for the generation and manipulation of abelian and non-abelian anyons \cite{Kitaev_2006,gohlke2018dynamical}. These have the form of flux-like excitations (also called vortices or visons) of the effective gauge field that emerges from the fractionalization of spin variables at each lattice site. Such excitations are conserved in the absence of external magnetic fields. 

%Here, we study the generation of such fluxes (also called vortices or visons) via the application of a local magnetic field in two distinct phases of the Kitaev model. We find that in both of them it is possible to generate and displace vortices with a reasonably high fidelity. 
%Thus, our work paves the way to the development of algorithm to perform quantum computation in topologically-ordered systems where visons can be generated and displaced via local scanning techniques. 

%Recently, neutron scattering revealed possible Kitaev quantum spin liquid phases in RuCl$_3$. The observation of a robust quantum thermal hall effect seems to corroborate this picture. Thus, it is natural to considers the possibility of performing quantum computation in such solid state systems. In this paper, we study protocols in which we apply the simplest possible perturbations: local and time-dependent magnetic fields, which can be generated with magnetic local scanning probes, to a sequence of atomic sites. We study two distinct phases of the Kitaev model, {\it i.e.} the $A_z$ and $B$ phases. We find that, in both phases, it is possible to generate and displace visons with a reasonably high fidelity. However the accumulated intrinsic errors of the protocol become eventually too large to perform braiding on scales large enough to avoid the mixing of the involved anyons. 

We propose a protocol in which, by adiabatically turning off the interactions between one spin and the rest of the system, and applying a local magnetic field on it, we can in principle displace visons and the attached Majorana zero mode with low errors.
The main target systems are quantum emulators~\cite{acharya2024quantumerrorcorrectionsurface} such as Rydberg atoms trapped in optical lattices \cite{ebadi2021quantum,morgado2021quantum}, that have recently emerged as a new platform to simulate strongly correlated systems
including the Kitaev model \cite{PhysRevResearch.6.L042054}.
Quantum emulators are composed of several qubits 
and allow for a great degree of control of the interactions between them. This in turn enables the realization of the Kitaev model and of the operations required to braid the anyons.

{\it The model}---We consider vison displacement in an extended Kitaev quantum spin liquid~\cite{Kitaev_2006} in the presence of a Zeeman field. By defining spin operators ${\sigma}_j$ in terms of the Majorana operators $b^x_j$, $b^y_j$, $b^z_j$ and $c_j$ at each lattice site $j$, the Hamiltonian of the system becomes ${\hat H}={\hat H}_0+{\hat H}_h$, where 
\begin{equation} \label{eq:Kitaev_Hamiltonian_Majorana}
    {\hat H}_{0} = -i \sum_{jk} J_{\alpha_{jk}} u_{jk} c_j c_k + \kappa\sum_{ijk} i u_{ij} D_j u_{jk}c_ic_k,
\end{equation}
is the (exactly solvable~\cite{Kitaev_2006}) extended Kitaev Hamiltonian, and ${\hat H}_h = \sum_{j} \bm{h}_j\cdot\bm{\sigma}_j$ is the Zeeman coupling to the external magnetic field. The gauge transformation operator $D_j= b^x_jb^y_jb^z_j c_j$ satisfies $D_j=1$ for physical states and $D_j=-1$ for unphysical ones~\cite{Kitaev_2006}. We note that $u_{jk}=i b^{\alpha_{jk}}_jb^{\alpha_{jk}}_k$, where $\alpha_{jk}$ takes the values $x,y,z$ depending on the direction of the bond $\langle j,k\rangle$, commute with ${\hat H}_{0}$ and are therefore constants of motions of the unperturbed extended Kitaev model. Their eigenvalues are $\pm 1$. The Hamiltonian~(\ref{eq:Kitaev_Hamiltonian_Majorana}) therefore describes the time evolution of the ``matter-like'' Majorana fermions $c$ in the presence of a $\mathbb{Z}_2$ gauge potential. To keep the model as general as possible, the magnitude of the Ising-like coupling $J_{\alpha_{jk}}$ is also allowed to depend on the bond direction. 
%The fermions acquire non-Abelian anyonic statistics when $\kappa\neq 0$. 
In solid state systems\cite{Kitaev_2006} the last term of~(\ref{eq:Kitaev_Hamiltonian_Majorana}) can arise from a magnetic field perturbation with non-zero components in all three $x,y,z$ directions.
Hereafter, we scale energies with $J$ (defined below) and we set $\hbar=1$. 

The original Kitaev Hamiltonian~\cite{Kitaev_2006} possesses a set of conserved quantities $W_p$, each defined on a hexagonal plaquette $p$. In the Majorana representation they are
$W_p=\prod_{\langle j,k\rangle}u^{\alpha_{jk}}_{jk}$,
where $\langle j,k\rangle$ cycles through the six sides of the hexagonal plaquette $p$ in the counterclockwise direction. Owing to Lieb's theorem~\cite{lieb1994flux}, the ground state of the Kitaev model must satisfy $W_p=1$ for any plaquette $p$. This state is said to be ``flux free''. In this paper, such state is defined by setting $u_{ij}^{\alpha_{ij}}=-1$ for $i$ ($j$) belonging to the $A$ ($B$) sublattice. When $W_p=-1$, the plaquette $p$ is said to be threaded by one flux (or vison).
%% Next we solve the three-spin Hamiltonian $\hat{H}_\kappa$. 
%%By the spin-to-Majorana mapping introduced above we also have 
%After inserting $D_j$ at the middle spin j we get
%% \begin{equation}
%% 	\hat{H}_\kappa=\sum_{ijk} i b_i^{\alpha_{ij}} b_j^\beta b_k^{\alpha_{jk}} c_ic_jc_k
%% \end{equation}
%% To convert it into a quadratic Hamiltonian in the $c$'s we multiply each term in it by a gauge transformation operator $D_j$, with $j$ being the middle spin. Physically relevant quantities are not affected by this operation because it preserves the physical states. After the transformation we have
%\begin{align}
%	\hat{H}_\kappa
% %    &=\kappa\sum_{ijk} i b_i^{\alpha_{ij}} b_j^\beta b_k^{\alpha_{jk}} c_ic_jc_k b_j^{\alpha_{ij}}b_j^{\alpha_{jk}}b_j^\beta c_j \\
%	% &=
%	% \kappa\sum_{ijk} -i b_i^{\alpha_{ij}} b_j^{\alpha_{ij}} b_j^{\alpha_{jk}} b_k^{\alpha_{jk}}c_ic_k\\
%	&=
%	\kappa\sum_{ijk} i u_{ij} D_j u_{jk}c_ic_k,
%\end{align}
%%
%therefore we effectively have a next-nearest-neighbour interaction. In the flux-free sector, this interaction opens up a gap between the matter-fermion bands, which become topologically non-trivial. 
When $\kappa\neq 0$, the fluxes carry a Majorana zero mode~\cite{Kitaev_2006} and have therefore non-Abelian statistics. 
This means that when $W_p = W_{p'} = -1$ in two far-apart plaquettes $p$ ad $p'$, the eigenstates of the Hamiltonian $\hat{H}_0$ split into a zero energy subspace and bands of excited states. The effective Hamiltonian in the zero energy subspace takes the forms $\hat{H}_{\text {eff}}=i\epsilon \hat{\gamma}_1\hat{\gamma}_2$ where $\hat{\gamma}_{1,2}$ are Majorana operators localized around plaquettes $p$ and $p'$. The exponentially small interaction energy is $\epsilon\propto e^{-L/\xi}$, where $L$ is the distance between the plaquettes and $\xi$ is the localization length. This generates the time evolution
\begin{equation}
    \hat{\gamma}_{1,2}(t)=\hat{\gamma}_{1,2} \cos(2 \epsilon t) \pm \hat{\gamma}_{2,1}\sin(2 \epsilon t)
\end{equation}
which implies that, on a time scale of order $1/\epsilon$, $\hat{\gamma}_1$ and $\hat{\gamma}_2$ exchange. Therefore, any quantum information stored in these qubits is protected only on a time scale of order $1/\epsilon$. We call this the ``dynamical error''. It follows that any braiding protocol must be executed faster than this time scale.

Another source of error comes from non-adiabatic transport. Suppose that at time $t=0$ there are two zero modes $\hat{\gamma}_1$ and $\hat{\gamma}_2$ separated by a distance $L$. After time $T$, the flux at 2 is brought to an adjacent plaquette 3, and the zero mode there is $\hat{\gamma}_3$. Now, because any realistic protocol can only be adiabatic to a certain degree, we have
\begin{align}
    \hat{U}(T)^\dagger \hat{\gamma}_1 \hat{U}(T) &=p_1\hat{\gamma}_1 + q_1\hat{\gamma}_3 +\hat{\delta}_1 \\
    \hat{U}(T)^\dagger \hat{\gamma}_2 \hat{U}(T) &=p_2\hat{\gamma}_3 + q_2\hat{\gamma}_1 +\hat{\delta}_2
\end{align}
where $\hat{U}(T)$ is the time evolution operator that takes the system through the protocol, $q_1$ and $q_2$ result from dynamical error, and $\hat{\delta}_{1,2}$ are the errors due to non-adiabaticity. We call $|p_{2}|^2$ the {\it Majorana transport probability}, and for protocols that are close to adiabatic, they are expected to be close to one. The parameter $q_1$ and $q_2$ can be made small by increasing the separation $L$, while $\hat{\delta}_1$ and $\hat{\delta}_2$ can be improved only by approaching the adiabatic limit.

The magnetic field ${\bm h}$ breaks the exact solvability of the model.
%, {\it i.e.} the ground state of ${\hat H}$ needs not to satisfy $W_p = 1$ for all $p$. 
Thus, to determine the time evolution, 
%under~(\ref{eq:driving_Hamiltonian}), 
we resort to a mean field approximation of ${\hat H}_{0}$. 
%For each bond, 
We decouple the four- and six-operator terms into Majorana bilinears 
%in Eq.~(\ref{eq:Kitaev_Hamiltonian_Majorana}) 
using Wick's theorem~\cite{minakawa2020majorana}.
%(schematically) as $b_A^\alpha b_B^\alpha c_Ac_B=\braket{b_A^\alpha b_B^\alpha }c_Ac_B
%    -\braket{b_A^\alpha c_A}b_B^\alpha c_B
%    -\braket{b_B^\alpha c_B}b_A^\alpha c_A
%    +\braket{b_A^\alpha c_B}b_B^\alpha c_A
%    +\braket{b_B^\alpha c_A}b_A^\alpha c_B$, where we used the sublattice indices $A,B$ to denote the two sites of a given bond. 
%Our treatment is similar to \cite{minakawa2020majorana} 
%Here, $\alpha$ is the component of the spin that is coupled along the bond. 
In this way, the Hamiltonian ${\hat H}_{0}$ is reduced to a quadratic form in the Majorana fermions, ${\hat H}_{MF}=\sum H_{MF}^{ij}m_i m_j$, where the matrix $H_{MF}$ depends on the values of mean fields~\footnote{ 
%as well as on the original Hamiltonian. 
Note that our mean field treatment is done in a single chosen gauge, and different gauges produce same physical results, just as in the normal Kitaev model.}. Hereafter, to make our notation more compact, we denote with $\{m_j\}$ the set of all $b_i^\gamma$ and $c_i$ Majorana fermions.
% (note that $j$ here does not denote lattice sites as before). 
% {\bf {\color{blue} Because we don't have the small h, we don't need to talk about how to diagonalize $H_{MF}$. {\color{red} (???)}} 
We evolve the single particle densities $\braket{m_i m_j}$ by solving the Heisenberg equation: 
    \begin{equation}
        \partial_t \braket{m_i(t) m_j(t)}=i[\hat{H}_{MF}(t),m_i(t) m_j(t)]
    \end{equation}
    At each time step, this equation gives a new value of $\braket{m_i(t)m_j(t)}$, which we then plug into $\hat{H}_{MF}$ to get the Hamiltonian for the next time step.

\begin{figure}[t]
    \centering
    \begin{tabular}{cc}
		\multicolumn{2}{c}{
		\begin{overpic}[width=0.9\linewidth]
            {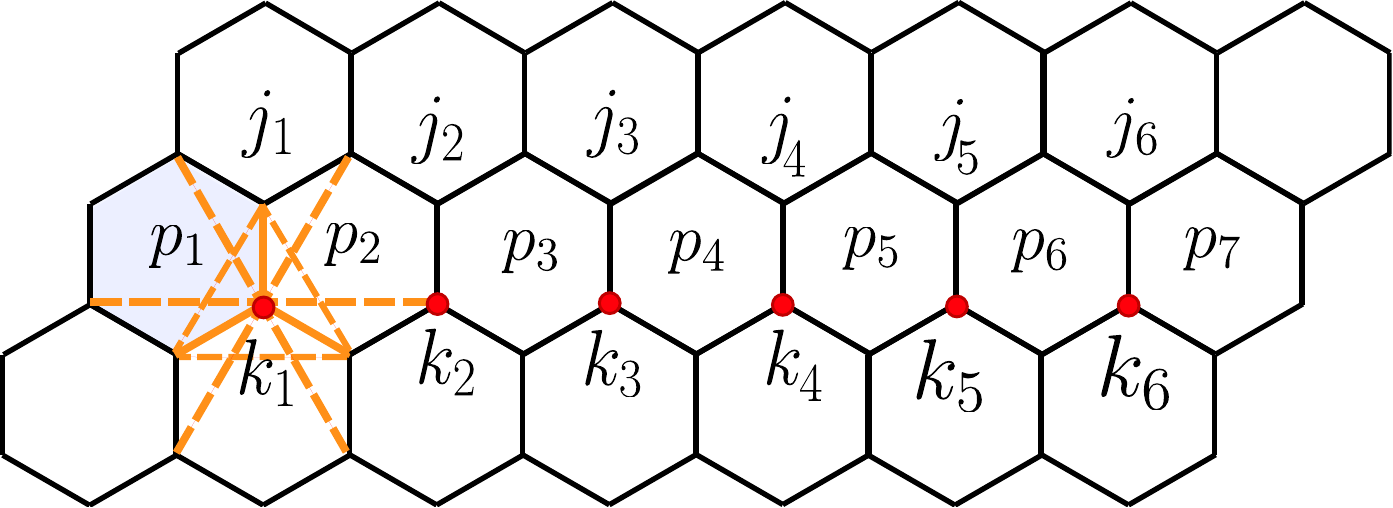}
            \put(0,-10){(a)}
        \end{overpic}
		}
		\vspace{0.5cm}\\
		\begin{overpic}[width=0.8\linewidth]
            {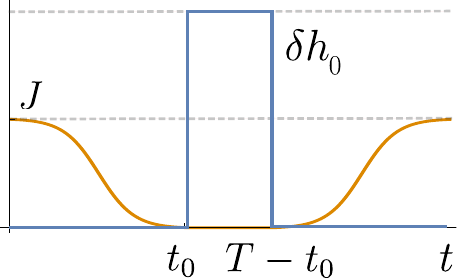}
            \put(0,-10){(b)}
        \end{overpic}
        % &
        % \begin{overpic}[width=0.49\linewidth]
        %     {fig14c.pdf}
        %     \put(0,-10){(c)}
        % \end{overpic}
    \end{tabular}
    \caption{(a) Illustration of the Kitaev model and of the flux manipulation protocol. Red circles are sites of the $A$ sublattice, denoted as $k_\ell$. Nearest-neighbour couplings are shown in black, and next-nearest-neighbour couplings are shown in dashed lines. The blue shaded plaquette $p_1$ stands for a flux, and at $t=0$, a time-dependent magnetic field in the $z$ direction is applied at site $k_1$. Its time-dependence follows the blue curve in (b). The orange shaded bonds (\textit{i.e.}, all bonds containing operators that live on $k_1$)
    % , which are all the bonds connected to $k_1$ (both NN and NNN couplings) but excluding the $j_1k_1$ bond, 
    are turned off between $0<t<t_0$ and back on between $T-t_0<t<T$ according to the orange curve in (b).%  We repeat this procedure on the sites $j_\ell$ and $k_\ell$, for $\ell=2,\ldots,6$. At the end of the protocol, the flux at $p_1$ is moved to $p_7$ {\bf (This is not really done in the main text?)}. %{\bf thicker lines, larger labels. Darken grey of bonds (maybe $B_\ell$ in white)}}
	(b) Time evolution of the magnetic field $\delta h$ (blue) and of the bond strengths (orange).
	% (c) Comparison of $u_1$ between two values of the turn-on-time $t_0$. We choose $T=10$, and use $\delta h=0.31J$ for $t_0=0$ and $\delta h=0.33J$ for $t_0=5J^-1$ (these values are chosen as to maximize the final values of $u$ reached). A finite value of $t_0$ breaks the exact solvability of the protocol, and the final value of $u_1$ reached is no longer exactly one. {\bf (legend is too small)} 
	}
    \label{fig:part2_protocol}
\end{figure}

{\it The protocol}---We now present a protocol which can displace Majorana zero modes with high probability. The protocol requires turning on and off interaction terms in $\hat{H}_0$, which is impossible in a solid state system, but can be realized in quantum emulators~\cite{acharya2024quantumerrorcorrectionsurface,ebadi2021quantum,morgado2021quantum}.
% Recent advances in developing emulators of surface codes open the possibility to remove restrictions existing in solid-state environments~\cite{acharya2024quantumerrorcorrectionsurface,ebadi2021quantum,morgado2021quantum}. %{\bf (add citation to arXiv:2408.13687)}.
%
%Our protocol goes as follows. Suppose 
We consider moving a vison from plaquette $p_1$ to $p_2$ of Fig~\ref{fig:part2_protocol}(a), thus simultaneously transporting a bound Majorana fermion when $\kappa\neq 0$. In the first step, 
we slowly 
%isolate the spin $k_1$, {\it i.e.}
%we slowly 
turn off all couplings between $k_1$ and neighbouring spins [orange in Fig~\ref{fig:part2_protocol}(a)].
The bonds are turned off between $0<t<t_0$ according to the curve~\footnote{
In our practical implementation, the all the couplings depend on time as
\begin{equation}
    f_{t_0}(t)=\frac{a}{1+e^{5-10t/t_0}}+b
\end{equation}
where $a,b$ are chosen such that $f_{t_0}(0)=1$ and $f_{t_0}(t_0)=0$.}
shown in yellow in Fig~\ref{fig:part2_protocol}(b).
Next, after the bonds are completely turned off and the spin is completely isolated, we apply a magnetic field $\delta h(t) \sigma_{k_1}^z$ to the spin $k_1$.
This generates the time evolution
%\begin{align}
%    \partial_t \braket{b_{k_1}^z b_{j_1}^z}&=-2\delta h(t) \braket{b_{j_1}^z c_1} \\
%    \partial_t \braket{b_{j_1}^z c_{k_1}}&=2\delta h(t) \braket{b_{k_1}^z b_{j_1}^z}
%\end{align}
%These equation take a closed form because we turned off the interactions between spin $k_1$ and the rest of the system. The solution takes the form
%
\begin{equation}
    \braket{b_{k_1}^z b_{j_1}^z}_t=\braket{b_{k_1}^z b_{j_1}^z}_{t_0}\cos\int_{t_0}^t 2\delta h(\tau)d\tau
    \label{eq:integral}
\end{equation}
where we used the condition $\braket{b_{j_1}^zc_{k_1}}_{t_0}=0$. This result is exact because we isolated $k_1$ from the rest of the system.
%The value of $u_1=\braket{ib_{k_1}^zb_{j_1}^z}$ is flipped when the integral $\int_{t_0}^t 2\delta h(\tau)d\tau=\pi$. 
In the remaining part of this paper we will assume that $\delta h(t)=\delta h_0$ for $t_0<t<T-t_0$ (where we choose the total time $T=2t_0+5J^{-1}$) and zero otherwise, {\it i.e.}, the magnetic field perturbation is turned on and off infinitely fast as shown by the blue curve in Fig~\ref{fig:part2_protocol}(b). 
%We use
Choosing $\delta h_0=\pi/[2(T-2t_0)]$
%
%\begin{equation}
%    \delta h_0=\frac{\pi}{2(T-2t_0)}
%\end{equation}
%
%for the strength of the magnetic field perturbation, which 
ensures that at time $t=T-t_0$, the value of $\braket{b_1^z b_2^z}$ is flipped~\footnote{Note that it is not necessary for $\delta h(t)$ to be constant. The bond value is flipped as long as the integral in Eqn~(\ref{eq:integral}) equals $\pi$ at time $t=T-t_0$.}. 
In other words, the magnetic field results in the unitary operator $\exp(i\pi \sigma^z_{k_1}/2)=i\sigma^z_{k_1}$ being applied to the spin $k_1$.
% and the particular shape of the magnetic field is irrelevant here.
%
Lastly, for the final step we slowly reconnect the spin to the rest of the system by bringing the couplings [orange in Fig~\ref{fig:part2_protocol}(a)] back to their original values over the time $T-t_0<t<T$. The flux and its associated bound zero mode have this been transported from plaquette $p_1$ to $p_2$. This procedure can be repeated to move the Majorana zero mode to any desired location. 

Note that the time evolution in Eq.~(\ref{eq:integral}) involves the mean field $\braket{b_{k_1}^z b_{j_1}^z}$ defined on the bond between spins $k_1$ and $j_1$. Such mean field is not bound to necessarily vanish because, although the bond has been cut off, the two spins can still be entangled. However, requiring them to be entangled puts a constraint on the time $T$ taken by the protocol to run. In fact, $T$ must be shorter than the typical times it takes for entanglement to decay (e.g., shorter than the decoherence times of the emulator).
Typical optical tweezer pick up times are on the order of $1~{\rm ms}$ \cite{barredo2016atom}. The decoherence times, on the other hand, depends on the type of system. 
State-of-the-art hyperfine qubits \cite{manetsch2024tweezer} can achieve decoherence times on the order of $10~{\rm s}$, while Rydberg atom arrays have decoherence time on the order of $1~{\rm ms}$ \cite{holzl2024long}. Therefore, realistic implementation of our protocol is feasible, albeit challenging in certain types of systems.

\begin{figure}[t]
    \centering
    \begin{tabular}{c}
        \begin{overpic}[width=0.9\linewidth]
            {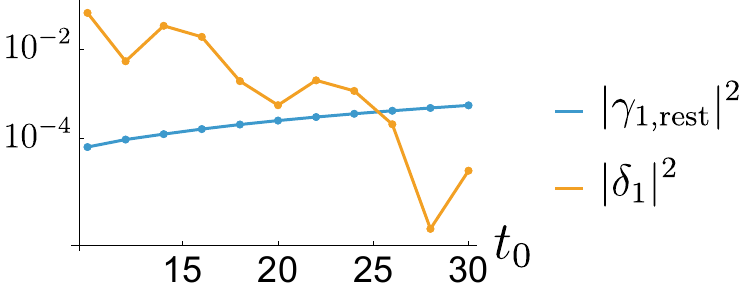}
            \put(0,-10){(a)}
        \end{overpic}
        \\
		\begin{overpic}[width=0.9\linewidth]
            {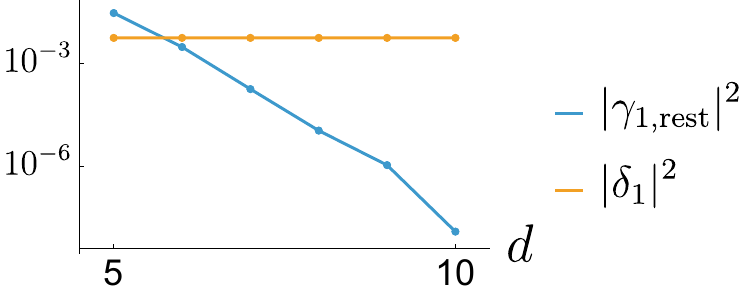}
            \put(0,-10){(b)}
        \end{overpic}
	%&
    %    \begin{overpic}[width=0.49\linewidth]
    %        {fig3d}
    %        \put(0,-10){(d)}
    %    \end{overpic}
    \end{tabular}
    \caption{(a) Dynamical and non-adiabatic errors as a function of the switching-on time $t_0$. The non-adiabatic error $|\delta_1|^2$ decays exponentially, while the dynamical error rises slowly. (b) Interaction and non-adiabatic errors as a function of the distance between fluxes $\gamma_1$ and $\gamma_2$. The dynamical error decays exponentially with distance. 
	% (b) Majorana transport probability $p_l$ over $l$ steps for four values of $T$. {\bf (larger legend)}
	}
    \label{fig:optimal_t0}
\end{figure}

To determine the error rate of our protocol, we apply it to a periodic system with $8\times20$ unit cells. Throughout this paper, we consider the system in the ferromagnetic phase and at the isotropic point, $J_x=J_y=J_z=-J$. We fix $\kappa=0.2J$.
We put a pair of fluxes at unit cells $(1,1)$ and $(5,10)$. We call the Majorana zero mode they carry $\hat{\gamma}_1$ and $\hat{\gamma}_{2}$, and their wave functions $\gamma_1$ and $\gamma_{2}$, respectively. We then move $\gamma_1$ by one unit cell to $(2,1)$ while keeping $\gamma_{2}$ stationary. For comparison, we also calculate the wave function of the Majorana zero mode $\beta_1,\beta_2$ localized at $(2,1),(5,10)$ by diagonalizing the Hamiltonian with fluxes located at $(2,1)$ and $(5,10)$. If transport were perfect, we would get $\gamma_1(T)=\beta_1$. 

In general, at the end of the protocol, the wave function $\gamma_1$ becomes
\begin{equation}
    \gamma_1(T)=\alpha \beta_1+ \gamma_{1,\text{rest}} +\delta_1
\end{equation}
where $\alpha\beta_1$ is projection of $\gamma_1(T)$ onto $\beta_1$, while $\gamma_{1,\text{rest}}$ is the projection of $\gamma_1$ onto $\beta_2$. %{\color{red} which remains practically unaltered by the protocol}. 
Finally, $\delta_1$ is the component of $\gamma_1(T)$ outside the zero energy subspace. %, which means the space spanned by the modes $g_1$ and $\gamma_2$. 
In Fig~\ref{fig:optimal_t0}(a) we show the {squared moduli of the wave functions} $|\gamma_{1,\text{rest}}|^2$ and $|\delta_1|^2$ as a function of $t_0$. %Hereafter, to shorten our notation, we use $|\psi|^2$ to denote the sum of the moduli squared of all the components of the wave function $\psi$. 
We find that the dynamical error $|\gamma_{1,\text{rest}}|^2$ slowly increases with $t_0$, as expected since the total time taken by the protocol $T=2t_0+5J^{-1}$ increases, and the dynamical phase scales as $T\epsilon$, where $\epsilon$ is the interaction energy between $\gamma_1$ and $\gamma_2$. Conversely, the non-adiabatic error $|\delta_1|^2$ shows an oscillating exponential decrease with $t_0$. Thus, as the time evolution becomes more adiabatic, the probability for the zero mode of remaining in the low-energy subspace increases.

In Fig~\ref{fig:optimal_t0}(b), we calculate the error rates for a system with $8\times30$ unit cells as a function of the distance between $\gamma_1$ and $\gamma_2$. We place the mobile zero mode $\hat{\gamma}_1$ at unit cell $(1,1)$ and the stationary mode $\hat{\gamma}_2$ at unit cell $(d,1)$. We move $\gamma_1$ by one unit cell to $(2,1)$, and calculate the error rates. In this calculation, we fix $t_0=12J^{-1}$. Fig~\ref{fig:optimal_t0}(b) shows that the dynamical error decays exponentially with $d$, while the non-adiabatic error barely changes. Thus, separating the two visons reduces their mixing during the time evolution.

\begin{figure}[t]
	\includegraphics[width=0.9\linewidth]{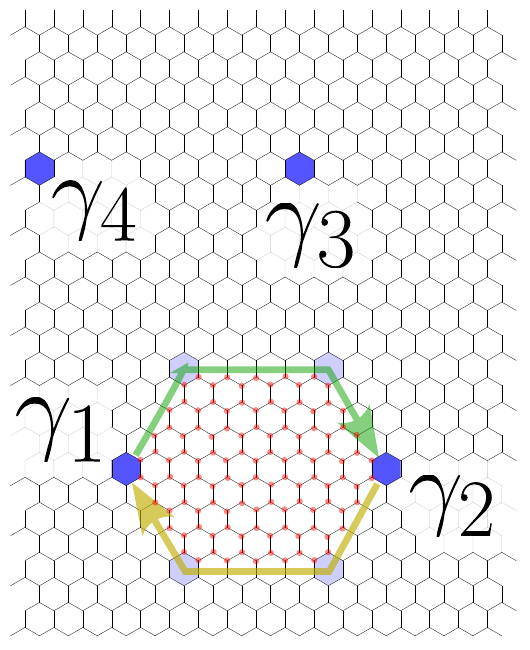}
	\caption{
%	Our periodic system has $15\times24$ unit cells. 
	At time $t=0$, four Majorana zero modes $\gamma_{i}$ ($i=1,\ldots, 4$) are placed in the four locations shown in this figure. We exchange $\gamma_1$ with $\gamma_2$ by moving $\gamma_1$ ($\gamma_2$) along the green (yellow) path, and then calculate their overlap with the initial zero modes. The sites shaded in red are the ones where a gauge transformation must be applied in order to meaningfully compare zero modes.}
	\label{fig:part2_braiding}
\end{figure}

% \subsection{Quantum gates}

{\it Braiding}---To demonstrate a simple quantum gate with our protocol, we consider a system of $15\times24$ unit cells and place four fluxes in the locations shown in Fig.~\ref{fig:part2_braiding}. We exchange two of them along the paths shown in the figure. %We set $\kappa=0.2J$, and put the system in the isotropic point of the ferromagnetic phase ($J_x=J_y=J_z=-J$).
%Unlike our previous protocol, we do not have a small magnetic field $h$ throughout the system.
We select two braiding times $t_0=20J^{-1}$ and $t_0=30J^{-1}$. 
%We then perform braiding with our protocol along the green and yellow paths in Fig.~\ref{fig:part2_braiding}, and 
We compare the modes $\gamma_{i}(t)$ ($i=1,\ldots,4$), obtained by braiding $\gamma_1$ and $\gamma_2$, with the modes $\beta_{i}(t)$.
%. The $\beta_i(t)$ modes are 
The latter are obtained by diagonalizing the Hamiltonian with the same gauge-field configuration produced by the braiding process at time $t$~\footnote{For instance, at the end of the protocol, the exchange operation flips the value of $u_{ij}$ along bonds crossing the edge of the hexagonal braiding region. This the final gauge configuration is different compared to the initial one (although the total flux enclosed by any loop is the same).}.
%Therefore, to compare the modes, we need to perform a gauge transformation}.

\begin{figure}[t]
	\centering
    \begin{tabular}{cc}
		\centering
        \begin{overpic}[width=0.5\linewidth]
            {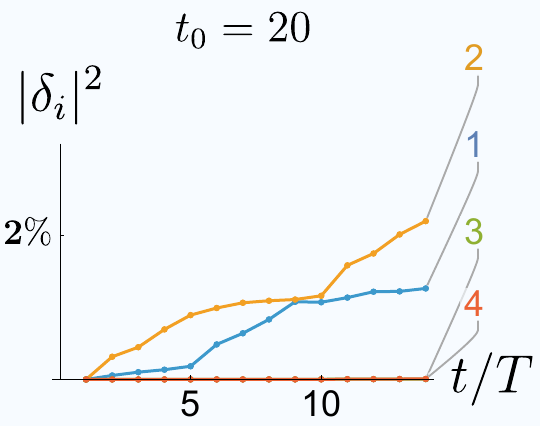}
            \put(0,-10){(a)}
        \end{overpic}
		&
		\begin{overpic}[width=0.5\linewidth]
            {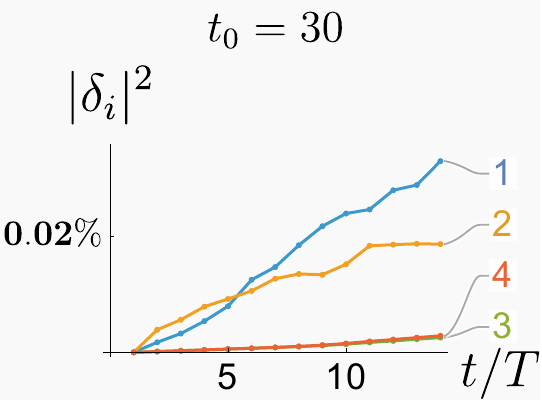}
            \put(0,-10){(b)}
        \end{overpic}
		\vspace{0.5cm}\\
		\centering
        \begin{overpic}[width=0.5\linewidth]
            {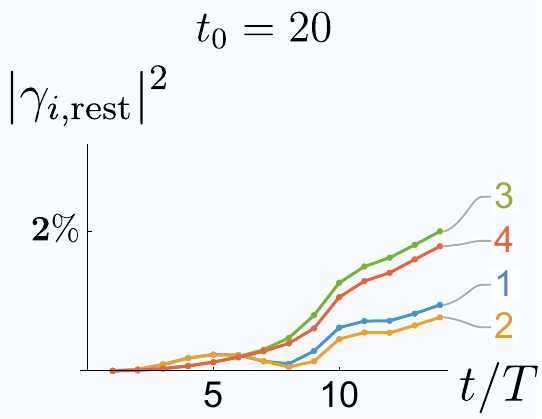}
            \put(0,-10){(c)}
        \end{overpic}
		&
		\begin{overpic}[width=0.5\linewidth]
            {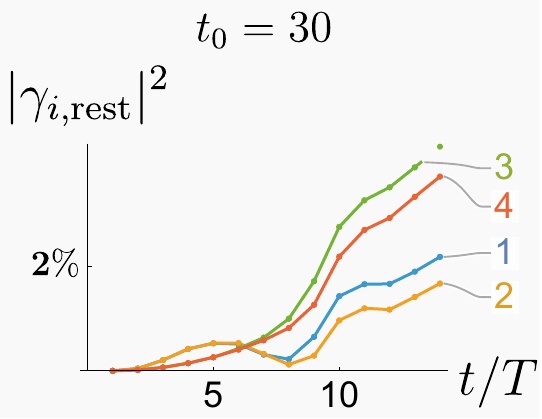}
            \put(0,-10){(d)}
        \end{overpic}
		\vspace{0.5cm}
		\\
		\centering
        \begin{overpic}[width=0.5\linewidth]
            {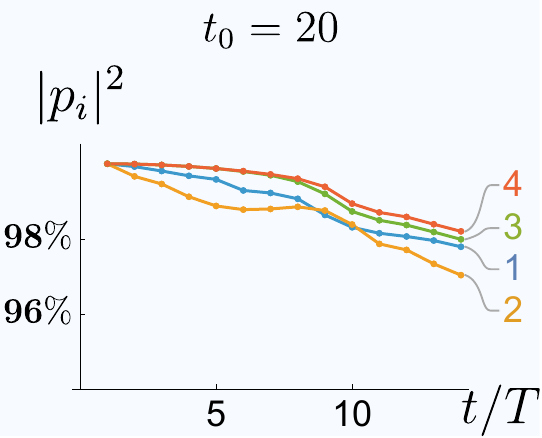}
            \put(0,-10){(e)}
        \end{overpic}
		&
		\begin{overpic}[width=0.5\linewidth]
            {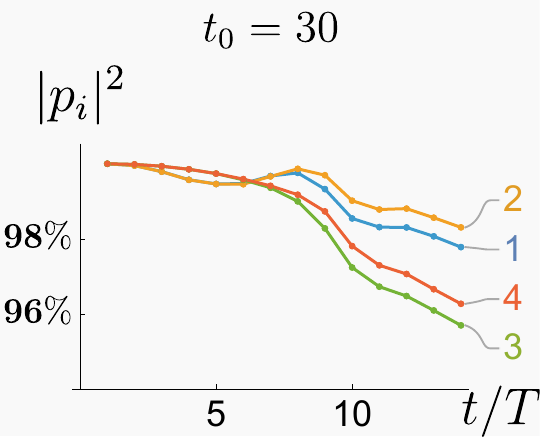}
            \put(0,-10){(f)}
        \end{overpic}
		\vspace{0.5cm}
	%&
    %    \begin{overpic}[width=0.49\linewidth]
    %        {fig3d}
    %        \put(0,-10){(d)}
    %    \end{overpic}
    \end{tabular}
    \caption{(a) and (b) Error $|\delta_i(t)|^2$ for Majorana fermion $i=1, \ldots 4$ due to non-adiabaiticity, plotted as a function of the braiding time $t$ for (a) $t_0=20J^{-1}$ and (b) $t_0=30J^{-1}$. Blue background stands for $t_0=20J^{-1}$ and grey for $t_0=30J^{-1}$. Time is measured in units of $T$, the time taken for the flux to move one unit cell. A larger $T$ is shown to greatly reduce non-adiabaticity errors.
	(c) and (d) Errors $|\gamma_{i,\text{rest}}|^2$ due to interactions between the fluxes. $t_0=30J^{-1}$ has larger errors because the dynamical phase is proportional to $\sim e^{iT_f\epsilon}$. (e) and (f) Success probability $p_i(t)$ as a function of time. 
    %Errors increase for stationary fluxes 3 and 4 after $t\sim6T$, which is due to small separation between the fluxes at this time.
	}
    \label{fig:part2_braiding_prob}
\end{figure}

The total running time is $T_f=13T$. In the ideal case, this operation would lead to~\footnote{Note that the signs here are arbitrary. We could define the Majorana modes as $\gamma_1'=-\gamma_1$, in which case $\gamma_1(T_f)=+\gamma_2$. If we fix the definition of $\gamma_1$ and $\gamma_2$, then the other sign choice corresponds to the counter-clockwise exchange of Majorana modes.} $\gamma_1(T_f)=-\gamma_2$ and $\gamma_2(T_f)=\gamma_1$,
%\begin{align}
%	\gamma_1(T_f)&=-\gamma_2 \\
%	\gamma_2(T_f)&=\gamma_1
%\end{align}
with $\gamma_{3}$ and $\gamma_4$ unchanged.  In other words, the quantum gate $\exp(-\pi\hat{\gamma}_1\hat{\gamma}_2/4)$
%\begin{equation}
%	\tau=e^{-\frac{\pi}{4}\gamma_1\gamma_2}
%	\label{eqn:idealmodes}
%\end{equation}
is applied. 
%Here, we write $\gamma_1$ for $\gamma_1(0)$ for simplicity.
%
We can thus write
\begin{equation}
	\gamma_i(t)=p_i(t)\beta_i(t)+\gamma_{i,\text{rest}}(t)+\delta_i(t)
\end{equation}
where $\beta_i(t)$ denotes the Majorana mode that $\gamma_i$ would evolve into if the braiding were perfect. 
$\gamma_{i,\text{rest}}(t)$ are again dynamical errors, {\it i.e.} undesired components of $\gamma_i(t)$ in the zero energy subspace.
%, {\bf {\it i.e.,} the components proportional to the other $\beta_j$'s where $j\neq i$}. 
Finally, $\delta_{i}(t)$'s are the non-adiabatic errors, {\it i.e.} the components outside the zero energy subspace.
In Fig.~\ref {fig:part2_braiding_prob}(a) and~(b) we plot the non-adiabatic errors $|\delta_i|^2$ for Majorana modes $i=1,\ldots, 4$ for $t_0=20J^{-1}$ and $t_0=30J^{-1}$, respectively. We find that the non-adiabatic errors remain close to zero for the stationary fluxes 3 and 4. This is expected, since the perturbation does not directly act on them. We also find that the non-adiabatic error for the mobile fluxes 1 and 2 for $t_0=30J^{-1}$ is lower by two orders of magnitude than for $t_0=20J^{-1}$. This is consistent with our previous result, which showed exponentially decaying non-adiabatic errors for larger values of $t_0$.

% For both $t_0$, we find that there are peaks in errors at times $t=5T$ and $t=10T$. We believe that these peaks are due to the fluxes making turns at the corners of the hexagon. We observe that at larger $T$ non-adiabatic errors are reduced for the mobile fluxes but are increased for the stationary fluxes. The reduction of non-adiabatic error for the mobile fluxes ($i=1,2$) at larger $T$ is expected, since larger $T$ leads to a better approximation of adiabatic transport. {\color{red} (what's the reason for larger $\delta_{3,4}$?)}
% {\color{red} (the following should be moved to the next paragraph)}
% Regarding the increase in $|\gamma_{i,\text{rest}}|^2$ error at larger $T$ for stationary fluxes ($i=3,4$), we believe that this is caused by the fact that they acquire a larger component proportional to mobile Majorana modes by interacting with them for a longer time. {\color{blue} Since the perturbation is acting directly on the mobile fluxes, the non-adiabatic error is larger for $T=200J^{-1}$ compared to $T=100J^{-1}$. 

% move a,b to c,d,}

In Fig.~\ref {fig:part2_braiding_prob}(c) and~(d) we plot the dynamical errors $|\gamma_{i,\text{rest}}|^2$ as a function of time. We find that the dynamical errors increase with $t_0$ (and therefore $T_f$). This is expected since the interaction between fluxes introduces a phase on the order $\sim e^{i T_f\epsilon}$ between zero-energy states.
% in the zero energy subspace.
%, and $T=2t_0+5J^{-1}=65J^{-1}$ is larger for $t_0=30J^{-1}$ than $T=\times45J^{-1}$ in the case of $t_0=20J^{-1}$. 
%For both times, 
We find that dynamical errors increase fast around $t\sim7T$. We attribute this behavior to the fluxes becoming insufficiently separated at that time, leading to a larger $\epsilon$ and faster error accumulation.
%than any other time. We believe that this behavior is due to the fluxes being insufficiently separated at time $t\sim7T$, leading to a larger $\epsilon$. 

In Fig.~\ref {fig:part2_braiding_prob}(e) and (f) we plot the probabilities $p_i(t)$. While trends are very similar, we find that probabilities improve for shorter switching-on times. This is due to the larger dynamical errors accumulated for longer braiding times {in our finite-size system}.
%, as non-adiabatic errors are negligibly small for $t_0=30$. 
For $t_0=20J^{-1}$, the total errors are larger for the mobile fluxes 1 and 2, and smaller for the stationary fluxes 3 and 4. This is because the non-adiabatic error is nearly zero for fluxes 3 and 4. On the other hand, the error for $t_0=30J^{-1}$ is dominated by the dynamical one.
Such errors can be exponentially suppressed by separating the Majorana zero modes by a large enough distance. 
% Therefore, it is reasonable to speculate that, for $t_0=30J^{-1}$ and for a larger system with longer distance between the fluxes the errors could be reduced. 
In turn, 
a larger system size 
% this 
would lead to improved results for $p_i$. 
{We note that, while making the system larger is theoretically possible, it may not be experimentally feasible. Hence, achieving a high probability of success of a quantum gate requires fine-tuning operation times.}

{\it Conclusions}---We have numerically studied the Kitaev model in the presence of local driving magnetic fields and tunable coupling strengths with a combination of mean-field and exact-diagonalization techniques. 
%We have shown that our method can reproduce some of the features found in numerical calculations performed with density-matrix renormalization group methods. In particular, it captures well the average value of the flux operator, a fact which lends credibility to our calculations of dynamical vortex generation.
%
%In this respect, w
% In the first part of this paper, we have studied the effect of a time-dependent and local driving magnetic field with a mean-field approach. We have demonstrated that such magnetic field can be used to create and move vortices in both the FM and AFM Kitaev model. We found the optimal parameters for such processes, and discussed the stability and quality of the fluxes produced. We also have shown that there is a one-to-one correspondence between the FM and AFM Kitaev models. We assessed the feasibility of our protocol in solid-state devices by modeling the effect of a local scanning tip. We found that the success probability is very low, even for the $A$ phase where a success probability $p_\text{flip}>90\%$ for single bonds was found. This is because the tip scans over a large region, which almost guarantees that lone fluxes are created. Thus, the protocol is doomed to fail somewhere along the way.
% {\color{red} (this part is not here anymore, correct? Also the abstract needs changing.)}
%
%
%In the second part of this paper, 
We presented a protocol for manipulating the Majorana zero modes, 
%obtained by applying both time-dependent magnetic fields and temporarily cutting off bonds around a given site. {\bf We 
discussing the error rates for a range of bond turn-off times $t_0$. We found that non-adiabatic transition amplitudes can be exponentially suppressed by increasing the time to switch off the interaction of a site with its neighbours. We then simulated a simple quantum gate by performing braiding on a system with four Majoranas and analysed the errors accumulated. We found that non-adiabatic transport and interaction between fluxes are the major sources of error. 
% Both of then can be 
These can be
exponentially suppressed by increasing the turn-off time and the distance between the fluxes.
{However, realistic implementations must consider the finite size of the system, while algorithms must run in a finite amount of time. Therefore, it is necessary to carefully consider the trade-off between different sources of error.} 

%while the latter can be reduced by placing the fluxes at longer distances apart, the former 
%appears unlikely to be overcome by increasing the time taken by the protocol. Therefore, in a realistic device, it is probably advisable to apply cooling measures which can relax the system to its ground state, which decreases the effects of non-adiabaticity in transport. 

% For a solid state device, our protocol is extremely challenging, as the operation of turning the bonds on and off is very difficult, if not impossible. 
%in a solid state device. 
% On the other hand, in emulators build with superconducting qubits or cold atoms, it might 
The protocol can be realized in emulators build with superconducting qubits or cold atoms.
There, interactions can be turned on and off,
for example slowly removing one of the atoms trapped in an optical lattice with optical tweezers \cite{anderegg2019optical,endres2016atom}.
% be possible to apply the equivalent of a magnetic field to a single atom and/or turn off bonds at will. 
% For example, one could imagine having atoms trapped in an optical lattice, and slowly pick up one of the atoms with an optical tweezer.
Provided that such operations are possible, and one can construct the Kitaev Hamiltonian \cite{will2025probing}, our protocol could make the braiding of non-abelian anyons possible.

{\it Acknowledgments}---We acknowledge support from the European Commission under the EU Horizon 2020 MSCA-RISE-2019 programme (project 873028 HYDROTRONICS) and from the Leverhulme Trust under the grant agreement RPG-2023-253.
\bibliographystyle{apsrev4-2}
\bibliography{biblio}
\end{document}